\documentclass[conference]{IEEEtran}
\IEEEoverridecommandlockouts

\usepackage{cite}
\usepackage{amsmath,amssymb,amsfonts}

\usepackage{algorithmic}
\usepackage{graphicx}
\usepackage{subcaption} 
\usepackage{float} 
\usepackage{textcomp}
\usepackage{xcolor}
\def\BibTeX{{\rm B\kern-.05em{\sc i\kern-.025em b}\kern-.08em
    T\kern-.1667em\lower.7ex\hbox{E}\kern-.125emX}}

\usepackage{tikz}
\usetikzlibrary{shapes.geometric, arrows, positioning}

\tikzstyle{input} = [ellipse, draw, text width=9em, text centered, minimum height=2em, font=\small, fill=none]
\tikzstyle{process} = [rectangle, draw, text width=10em, text centered, minimum height=2em, font=\small, fill=none]
\tikzstyle{output} = [rectangle, draw, text width=10em, text centered, minimum height=2em, font=\small, fill=none]
\tikzstyle{arrow} = [thick,->,>=stealth]

\begin{document}

\title{Knowledge Graph-Based Multi-Agent Path Planning in Dynamic Environments using WAITR}

\author{\IEEEauthorblockN{Ted Edward Holmberg}
\IEEEauthorblockA{\textit{Cannizaro-Livingston Gulf States} \\
\textit{Center for Environmental Informatics}\\
New Orleans, LA, USA \\
eholmber@uno.edu}
\and
\IEEEauthorblockN{Elias Ioup}
\IEEEauthorblockA{\textit{Center for Geospatial Sciences} \\
\textit{Naval Research Laboratory}\\
Stennis Space Center, Mississippi, USA \\
elias.z.ioup.civ@us.navy.mil}

\and
\IEEEauthorblockN{Mahdi Abdelguerfi}
\IEEEauthorblockA{\textit{Cannizaro-Livingston Gulf States} \\
\textit{Center for Environmental Informatics}\\
New Orleans, LA, USA \\
mahdi@cs.uno.edu}
}

\maketitle

\begin{abstract}
This paper addresses the challenge of multi-agent path planning for efficient data collection in dynamic, uncertain environments, exemplified by autonomous underwater vehicles (AUVs) navigating the Gulf of Mexico. Traditional greedy algorithms, though computationally efficient, often fall short in long-term planning due to their short-sighted nature, missing crucial data collection opportunities and increasing exposure to hazards. To address these limitations, we introduce WAITR (Weighted Aggregate Inter-Temporal Reward), a novel path-planning framework that integrates a knowledge graph with pathlet-based planning, segmenting the environment into dynamic, speed-adjusted sub-regions (pathlets). This structure enables coordinated, adaptive planning, as agents can operate within time-bound regions while dynamically responding to environmental changes. WAITR's cumulative scoring mechanism balances immediate data collection with long-term optimization of Points of Interest (POIs), ensuring safer navigation and comprehensive data coverage. Experimental results show that WAITR substantially improves POI coverage and reduces exposure to hazards, achieving up to 27.1\% greater event coverage than traditional greedy methods.
\end{abstract}

\begin{IEEEkeywords}
Dynamic Path Planning, Autonomous Underwater Vehicles, Knowledge Graphs, Points of Interest, Greedy Algorithms, Cumulative Scoring
\end{IEEEkeywords}

\section{Introduction}

Efficient navigation and data collection are essential in dynamic, uncertain environments, such as marine ecosystems or remote monitoring zones. Autonomous systems, including autonomous underwater vehicles (AUVs) operating in regions like the Gulf of Mexico, underscore the challenge of gathering spatially and temporally optimized environmental data \cite{rudnick2015ocean_research, chassignet2007hycom}. In these settings, Points of Interest (POIs) not only vary by location but also shift in significance over time due to changing factors such as water temperature, salinity, and current velocity \cite{holmberg2014data_visualization}.

Current path-planning methods, particularly those employing greedy algorithms, are widely favored for their simplicity and computational efficiency \cite{xu2020greedy_dynamic}. Greedy algorithms focus on immediate rewards, such as maximizing data collection at a POI or avoiding hazards \cite{mei2024multi_agent}. These approaches have proven effective in both single-agent and multi-agent systems for real-time sensor placement and environmental monitoring. Notable frameworks, including those by Mei et al. \cite{mei2024multi_agent} and Xu et al. \cite{xu2020greedy_dynamic}, highlight the utility of greedy approaches in dynamic environments. 

However, the short-term focus of greedy algorithms often results in suboptimal long-term outcomes, particularly in environments with evolving POIs and unpredictable conditions \cite{wu2014path_problems}. As environmental factors shift, agents using greedy algorithms may miss opportunities for high-value data collection or fail to effectively avoid hazards due to limited predictive capabilities. This paper introduces WAITR (Weighted Aggregate Inter-Temporal Reward), a prediction-driven path-planning framework that addresses these limitations. WAITR integrates a knowledge graph encoding spatial and temporal relationships with pathlet-based planning and cumulative scoring. By segmenting the environment into dynamic, speed-adjusted sub-regions (pathlets) and leveraging real-time updates, WAITR enables agents to anticipate environmental changes and adapt their paths adaptively \cite{lavalle2006planning}.

Designed to balance immediate rewards with long-term optimization, WAITR supports scalable, adaptive planning for multi-agent spatiotemporal path planning. By bridging the gap between short-term efficiency and robust, predictive optimization, WAITR offers a domain-agnostic solution that enhances data coverage and navigational safety across dynamic environments.

\section{Background}
\subsection{Spatiotemporal Points of Interest (POI)}
A Point of Interest (POI) is a location in both space and time where data collection or observation is required, with the specific criteria for identifying POIs defined by the practitioner \cite{rudnick2015ocean_research}. The parameter determining a POI is flexible, allowing the path-planning model to be adapted to various tasks and environments \cite{chassignet2007hycom}. The POI can represent any relevant parameter depending on the research objectives.  Systems like the Geospatial Information Distribution System (GIDS) \cite{chung2001gids} and research on spatiotemporal information systems \cite{ladner2012mining} have been developed to address challenges in distributing and integrating geospatial data from multiple sources.

In the case study presented in this paper, focusing on the Gulf of Mexico (GoM), POIs include regions where significant environmental data, such as chlorophyll levels, salinity, and temperature, needs to be collected \cite{chassignet2007hycom}. These regions are of both spatial and temporal significance, influenced by dynamic factors like temperature gradients, salinity shifts, and ocean currents. The changing nature of these environmental conditions necessitates adaptive monitoring strategies to ensure efficient and comprehensive data collection over time \cite{holmberg2022geo_spatiotemporal}.

\subsection{Dynamic Environmental Conditions}
In spatiotemporal path-planning systems, environmental conditions play a critical role in defining the ease of traversal through different terrains \cite{lavalle2006planning}. These conditions can vary widely, influenced by factors such as shifting water currents, muddy or flooded roads, or high winds \cite{thrun2005probabilistic}. The dynamic nature of these conditions directly impacts the energy, time, and safety costs for agents as they navigate through the environment.

For mobility-constrained agents such as low-powered underwater gliders, land-based vehicles on rough terrain, or drones operating in adverse weather conditions, environmental changes can significantly alter their ability to move efficiently \cite{rudnick2015ocean_research}. As these factors evolve over time, previously navigable areas may become more challenging or hazardous, requiring the system to adjust to these changing conditions. The need to account for these variations highlights the importance of dynamic path planning that responds to evolving terrain costs, ensuring that agents can continue to operate effectively across varying environmental scenarios \cite{xu2020greedy_dynamic}.

\subsection{Clustering Points of Interest (POIs)}
Efficient monitoring and data collection in dynamic environments rely on identifying and prioritizing critical regions known as Points of Interest (POIs). These POIs represent areas where observations are most valuable. The environment is divided into spatial grids, and Proximal Recurrence (PR) clustering is applied to detect regions of significant environmental activity \cite{holmberg2023stroobnet}. By considering both spatial and temporal dynamics, PR identifies areas where environmental conditions are evolving, flagging them as priority zones for autonomous agents \cite{wu2014path_problems}. This structured framework for POI identification forms the foundation for agent path planning, focusing attention on regions undergoing notable changes.

The PR algorithm, outlined by Holmberg et al. (2023) \cite{holmberg2023stroobnet}, effectively captures dynamic environmental patterns, allowing agents to adjust routes as conditions shift. The proposed approach builds on this foundation with the Weighted Proximal Recurrence (WPR) algorithm, which further refines path planning by weighting POIs according to potential data value, environmental risk, and prediction confidence. This enhancement allows agents to prioritize areas with high-value data and lower risk, optimizing data collection and navigational safety in spatiotemporal contexts.

\section{Related Work}

Path planning, a fundamental problem in robotics and artificial intelligence, seeks to find optimal or near-optimal paths for agents navigating through an environment \cite{lavalle2006planning}. Various approaches have been proposed to address this challenge, each with its own strengths and limitations. This section reviews relevant path planning techniques, focusing on those that leverage graph-based representations and address dynamic environments, and highlights the limitations that motivate our proposed approach.

\subsection{Graph-Based Path Planning}

Graph-based methods are widely used in path planning due to their ability to represent complex environments and efficiently find optimal paths \cite{lavalle2006planning, thrun2005probabilistic}. These methods model the environment as a graph, where nodes represent locations (e.g., cluster centers, Points of Interest (POIs)) and edges represent possible transitions between them. Edge weights can be assigned based on factors such as distance, traversal cost, or environmental hazards \cite{chassignet2007hycom}. Common graph-based algorithms include Dijkstra's algorithm and A*, which efficiently find the shortest path between a starting and goal node. 

While effective in static environments, traditional graph-based methods often face challenges in dynamic settings where environmental conditions change over time. For example, changes in water currents or the emergence of new obstacles can alter edge weights or node connectivity, rendering pre-computed paths suboptimal or even infeasible \cite{stentz1995optimal}. Addressing these dynamic aspects requires either frequent replanning or incorporating predictive models into the path planning process \cite{ferguson2006motion}. 

\subsection{Greedy Algorithms}

Greedy algorithms are another popular approach for path planning, particularly in dynamic environments, due to their computational efficiency and simplicity \cite{xu2020greedy_dynamic, mei2024multi_agent}. These algorithms make locally optimal decisions at each step, aiming to maximize immediate rewards without considering the long-term impact on the overall path \cite{wu2014path_problems}. For example, a greedy algorithm might prioritize visiting the nearest POI with the highest immediate data collection value, regardless of potential future gains or risks.

However, the inherent myopic nature of greedy algorithms often leads to suboptimal solutions in dynamic environments. As conditions evolve, agents guided by greedy decisions may miss opportunities to collect more valuable data in the future or navigate to safer regions \cite{xu2020greedy_dynamic, mei2024multi_agent}. This shortsightedness limits their effectiveness in scenarios where long-term planning and adaptation to changing conditions are crucial.

\subsection{Knowledge Graphs for Path Planning}

Recent research has explored the use of knowledge graphs to enhance path planning in complex and dynamic environments. Knowledge graphs are a powerful tool for representing semantic information about the environment, including spatial relationships, temporal dynamics, and agent capabilities \cite{chen2020knowledge}.  A knowledge graph can encode information about the likelihood of encountering hazards in different regions or the predicted changes in POI importance over time. By incorporating this knowledge into the path planning process, agents can make more informed decisions that consider both short-term gains and long-term consequences.

However, existing approaches that utilize knowledge graphs for path planning often focus on single-agent scenarios and do not explicitly address the challenges of multi-agent coordination in dynamic environments \cite{khan2018knowledge}. Furthermore, incorporating real-time updates and predictions into the knowledge graph and efficiently querying it for path planning decisions remain open challenges.

\subsection{Addressing the Limitations}

This paper addresses the limitations of existing path planning approaches by proposing a novel method that integrates a knowledge graph structure with pathlet-based planning and a cumulative scoring mechanism called WAITR. By leveraging the knowledge graph's ability to represent dynamic environmental information and predict future changes, WAITR enables agents to make more informed decisions that balance immediate gains with future potential. This approach aims to overcome the myopic nature of greedy algorithms and provide a scalable and robust solution for multi-agent path planning in complex, dynamic, and uncertain environments.

\section{Mathematical Foundations}

This section presents the mathematical foundations underlying our proposed approach for multi-agent path planning in dynamic environments. We introduce the Proximal Recurrent Event Partition (PREP) Mapper for identifying significant Points of Interest (POIs), the Temporal Event Dynamics (TED) Predictor for capturing dynamic POI shifts, and the Weighted Aggregate Inter-Temporal Reward (WAITR) Planner for optimizing agent paths.

\subsection{Proximal Recurrent Event Partition (PREP) Mapper}

The PREP Mapper aims to identify clusters of POIs that are both spatially and temporally significant. It operates on a graph representation of the environment, denoted by \(G = (V, E)\), where nodes \(V\) represent potential sensor locations and edges \(E\) represent possible paths between them. The PREP Mapper's objective function seeks to maximize the overall reward of visiting a set of POIs while minimizing the travel distance between them:

\[
\text{arg max}_{C} \sum_{i \in C} W(i) - \lambda \cdot D(i, i+1)
\]

Here, \(C\) represents a cluster of POIs, \(W(i)\) denotes the reward associated with visiting POI \(i\), \(D(i, i+1)\) represents the distance between consecutive POIs in the cluster, and \(\lambda\) is a balancing coefficient that controls the trade-off between reward maximization and distance minimization. 

Intuitively, the PREP Mapper seeks to find clusters of POIs that offer high rewards while being relatively close to each other, ensuring efficient data collection and minimizing travel costs. The balancing coefficient \(\lambda\) allows us to adjust the importance of these two objectives based on the specific application requirements.

\subsection{Temporal Event Dynamics (TED) Predictor}

The TED Predictor captures the dynamic nature of POIs by identifying when their significance changes over time. It analyzes the event density \(\rho(t)\), which represents the frequency of events occurring at a given location over time. The TED Predictor identifies a POI as dynamically significant if the rate of change in event density exceeds a certain threshold:

\[
\frac{\partial}{\partial t} \rho(t) > \delta
\]

where \(\frac{\partial}{\partial t} \rho(t)\) represents the rate of change of event density with respect to time, and \(\delta\) is a threshold that defines the minimum change required to trigger a path adjustment.

This criterion ensures that the TED Predictor accurately captures shifts in POI importance, enabling agents to adapt their paths proactively to target regions where significant events are likely to occur in the future. The threshold \(\delta\) can be adjusted based on the desired sensitivity to changes in event density.

\subsection{WAITR Planner Optimization Problem}

The WAITR Planner aims to optimize agent paths by maximizing the cumulative reward collected over time while considering future potential benefits. It formulates the path planning problem as a maximization problem over a set of waypoints \(P\) selected from the graph's nodes \(V\):

\[
\max_{P \subseteq V} \sum_{t=0}^{T} \gamma^t  \left( \sum_{i=1}^{|P_t|} W(P_{t,i}) - \lambda \cdot R(P_{t,i}, P_{t,i+1}) \right)
\]

In this formulation:

\begin{itemize}
    \item \(T\) represents the planning horizon, or the number of time steps considered in the optimization.
    \item \( \gamma \in (0, 1] \) is a discount factor that weights the importance of future rewards, allowing for a trade-off between immediate and future gains.
    \item \(P_t\) denotes the set of waypoints visited at time step \(t\).
    \item \(W(P_{t,i})\) represents the reward associated with visiting waypoint \(P_{t,i}\) at time \(t\), which can be influenced by predicted events or knowledge decay.
    \item \(R(P_{t,i}, P_{t,i+1})\) represents a penalty for the distance between consecutive waypoints at time \(t\).
    \item \(\lambda\) is a balancing coefficient that controls the trade-off between reward maximization and travel cost minimization.
\end{itemize}

The WAITR Planner uses a external predictor to estimate future rewards \(W(P_{t,i})\), taking into account potential events and knowledge decay. Uncertainty in future predictions can be incorporated by assigning confidence scores to the forecasted events or by adjusting the discount factor \(\gamma\) to prioritize more certain near-term rewards over less certain long-term rewards. Additionally, the impact of hazards can be computed into the reward function, either through the edge weights \(R(P_{t,i}, P_{t,i+1})\) or by directly penalizing paths that traverse high-risk regions.

\section{Methodology}

This section details our proposed approach for multi-agent path planning using the WAITR (Weighted Aggregate Inter-Temporal Reward) algorithm and its integration with a knowledge graph.

\subsection{Knowledge Graph Construction}

We begin by constructing a dynamic knowledge graph to represent the GoM environment. The graph encodes crucial information for path planning and consists of:

\begin{itemize}
    \item \textbf{Nodes:} Representing POIs, hazards (strong UV currents), and bridge points for navigation. Each node contains temporal information, reflecting changing environmental conditions (e.g., temperature, currents).
    \item \textbf{Edges:} Representing possible paths between nodes, encoded with dynamic weights that reflect traversal difficulty based on distance and potential hazards. 
\end{itemize}

This knowledge graph enables:

\begin{itemize}
    \item \textbf{Real-Time Adaptation:}  Nodes and edges are updated dynamically, enabling agents to adapt to changing conditions and newly discovered hazards.
    \item \textbf{Foresight:} Temporal edges encode predicted future changes in environmental conditions and POI importance, allowing agents to plan more strategically.
    \item \textbf{Hazard Mitigation:}  Edge weights guide agents to avoid risky areas, enhancing safety while achieving data collection objectives.
    \item \textbf{Multi-Agent Coordination:}  Shared knowledge within the graph facilitates agent coordination, minimizing path conflicts and optimizing overall coverage.
\end{itemize}

\begin{figure}[ht]
\centering
\includegraphics[width=0.75\columnwidth]{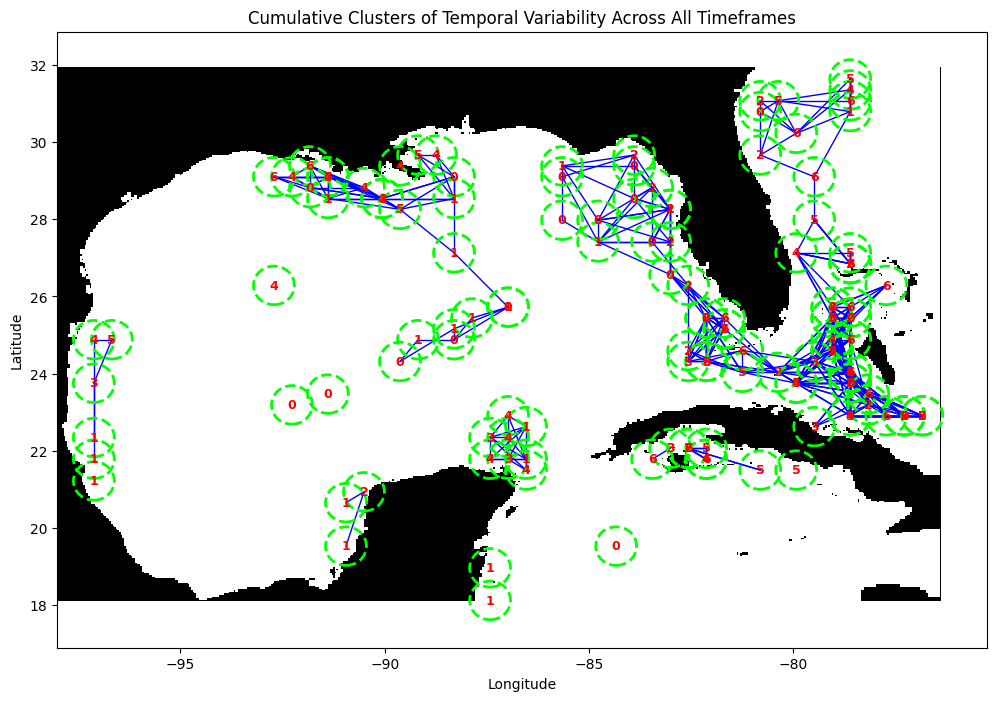} 
\caption{ROBUSTnet, Aggregated Spatiotemporal Graph}
\label{fig:robust-net} 
\end{figure}

\begin{figure}[ht]
    \centering
    \begin{tikzpicture}[node distance=2cm] 

        \node (input) [input] {Real-time Data \\ (Env. Conditions)};
        
        \node (kg) [process, below of=input] {Knowledge Graph \\ Nodes: POIs, Hazards \\ Edges: Relationships};
        
        \node (pathlet) [process, below of=kg] {Pathlet Selection \\ Organize Subgraphs \\ Efficient Decisions};
        
        \node (output) [output, below of=pathlet] {Optimized Path \\ (Path Planning)};
        
        \draw [arrow] (input) -- (kg);
        \draw [arrow] (kg) -- (pathlet);
        \draw [arrow] (pathlet) -- (output);
        
        \node (dynamic) [process, right of=kg, xshift=2cm] {Dynamic Updates \\ Hazards and POIs};
        \draw [arrow] (dynamic) -- (kg);
        
        \node (scoring) [process, right of=pathlet, xshift=2cm] {Cumulative Scoring \\ WAITR Algorithm};
        \draw [arrow] (scoring) -- (pathlet);

    \end{tikzpicture}
    \caption{Flowchart of the data processing pipeline incorporating the Knowledge Graph, Pathlet Selection, and Scoring mechanisms in the WAITR Algorithm.}
    \label{fig:waitr-flowchart}
\end{figure}
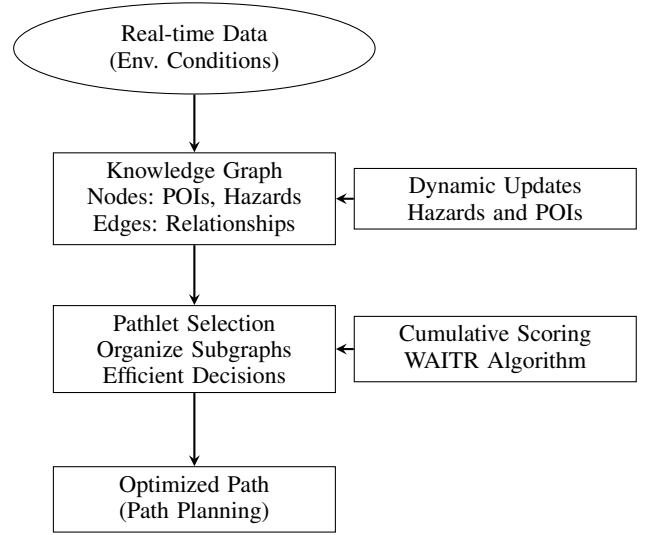

Our knowledge graph extends traditional path-planning methods. Figure \ref{fig:waitr-flowchart} illustrates the data processing pipeline incorporating the knowledge graph, pathlet selection, and scoring mechanisms used in the WAITR algorithm.  Figure \ref{fig:robust-net} visualizes an aggregated spatiotemporal graph, showcasing the interconnected nature of POIs, hazards, and potential paths.

\subsection{Benefits Over Traditional Graphs}

Traditional graphs typically capture only static spatial relationships, limiting their ability to adapt to environmental changes. In contrast, our knowledge graph supports real-time decision-making by dynamically updating nodes and edges to reflect shifting conditions. This dynamic nature enables agents to:

\begin{itemize}
    \item \textbf{Adapt to changing conditions:} Nodes and edges adjust in real time based on evolving hazards, offering advantages in responsiveness compared to static models \cite{holmberg2022geo_spatiotemporal}.
    \item \textbf{Plan with foresight:} Temporal edges allow agents to consider forecasted changes and adjust their paths proactively.
\end{itemize}

\subsection{Pathlet-Based Path Planning}

A \textit{pathlet} is a localized subgraph within a larger knowledge graph, representing a specific, bounded region of the environment. Pathlets serve to focus the decision-making process by limiting the scope to a manageable subset of nodes, which reduces computational complexity for large-scale path planning. By dividing the environment into pathlets, agents can optimize their movements locally within each pathlet while still maintaining the flexibility to transition between adjacent pathlets for broader, global exploration. 

Precomputed shortest paths within each pathlet are stored in lookup tables, allowing agents to make efficient real-time decisions without recalculating paths on the fly. This approach enhances scalability by balancing local optimization within pathlets with the potential for global navigation.

\subsection{WAITR Algorithm}

The WAITR algorithm guides agents to prioritize POIs and navigate the knowledge graph. It employs a cumulative scoring system that balances short-term gains with long-term rewards, considering:

\begin{itemize}
    \item \textbf{POI Importance:}  Rewards associated with visiting POIs, which can be influenced by predicted future events or knowledge decay.
    \item \textbf{Travel Costs:}  Penalties based on the distance between waypoints.
    \item \textbf{Hazard Levels:}  Penalties for traversing hazardous regions, integrated into the reward function through edge weights.
\end{itemize}

WAITR incorporates an external predictor (not discussed in detail here) to estimate future POI importance based on potential events and knowledge decay. It accounts for uncertainty in these predictions through confidence scores and a discount factor (\(\gamma\)), prioritizing near-term rewards over less certain long-term rewards.

\subsection{Weighted Clustering of Points of Interest (POIs)}

The initial waypoint placement within the knowledge graph is determined using the Weighted Proximal Recurrence (WPR) algorithm. WPR enhances the standard Proximal Recurrence (PR) method \cite{holmberg2023stroobnet} by incorporating environmental risk factors and the confidence in predicted data significance into the clustering process.  

In WPR, each POI is weighted based on the potential data yield, the calculated environmental risk, and the confidence in the predicted future value of the POI. This allows the system to prioritize regions that offer a good balance between high-value data, low risk, and high prediction confidence. The weighting function is represented as:

\[
W(i) = \alpha \times C(i) \times V(i) - \beta \times R(i)
\]

where:

\begin{itemize}
    \item \(V(i)\) represents the data value of POI \(i\).
    \item \(R(i)\) represents the associated environmental risk factor (e.g., strength of currents, presence of obstacles).
    \item \(C(i)\) represents the confidence score associated with the predicted future value of POI \(i\), ranging from 0 (no confidence) to 1 (full confidence).
    \item \(\alpha\) represents the risk tolerance coefficient, determining how much emphasis is placed on minimizing risk. A higher \(\alpha\) indicates a more risk-averse approach.
    \item \(\beta\) represents the confidence weighting coefficient, determining the influence of the prediction confidence on the overall weighting. A higher \(\beta\) indicates that the system prioritizes POIs with higher confidence scores.
\end{itemize}

\begin{figure}[htbp]
\centering
\includegraphics[width=0.75\linewidth]{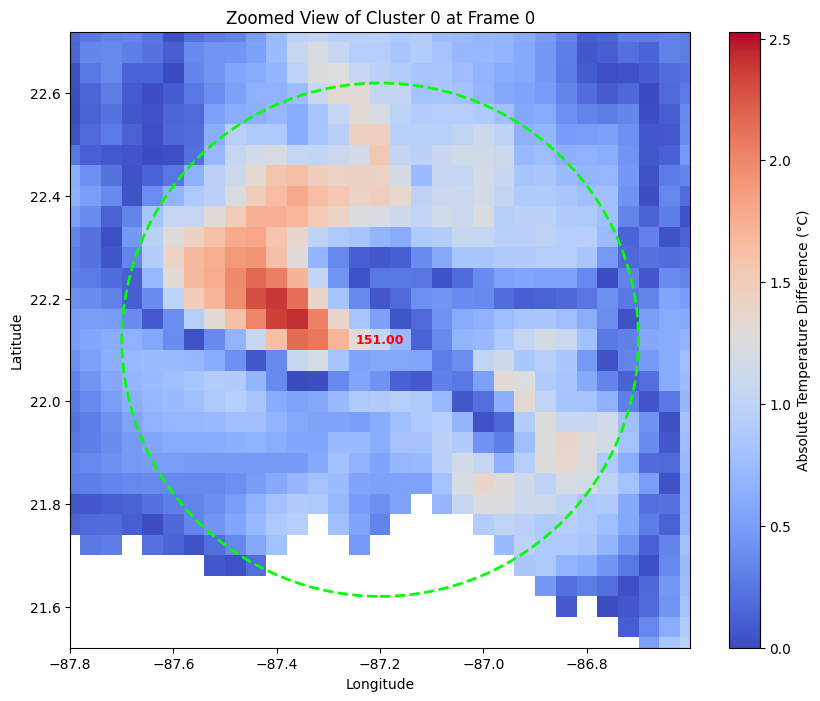}
\caption{Weighted observation radius with hazards factored in. The score for the observation centroid (151.00) highlights the balance between potential data collection and environmental risks.}
\label{fig:weighted-cluster}
\end{figure}

\begin{figure}[ht]
\centering
\includegraphics[width=0.75\columnwidth]{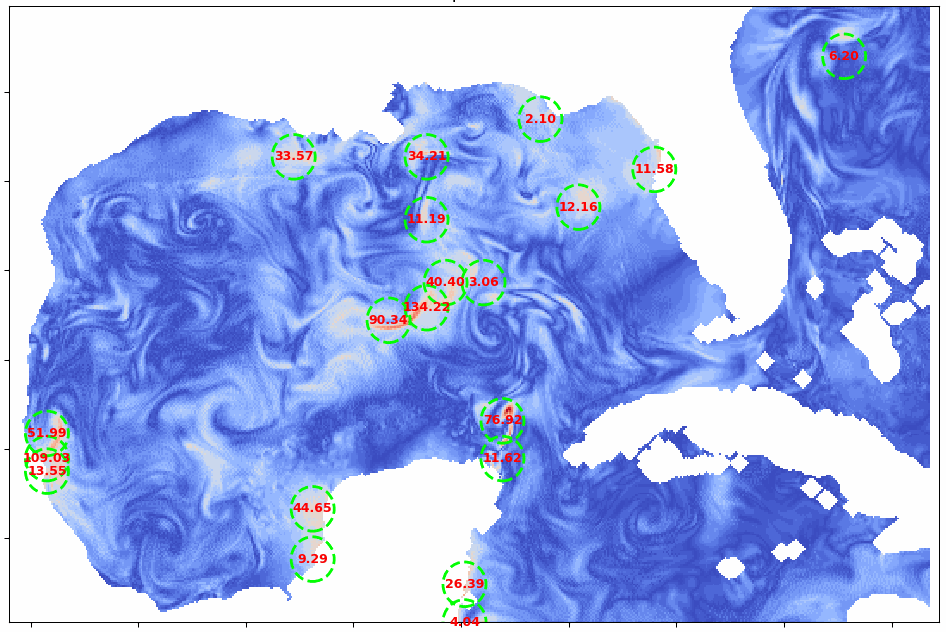} 
\caption{Shows densest cluster identifications for frame 1}
\label{fig:wpr} 
\end{figure}

Figures \ref{fig:weighted-cluster} and \ref{fig:wpr} illustrate the WPR process, showcasing how these factors are considered when determining initial waypoint placements.  

By incorporating risk tolerance and confidence weighting, WPR enables agents to make more informed path choices that consider both the immediate and predicted future conditions. This advancement in clustering optimizes the path-planning process in spatiotemporal contexts, ensuring that agents prioritize high-value, low-risk areas with reliable predictions as they adapt to the dynamic environment.

\subsection{Temporal Evolution of Clusters}

The clusters of POIs evolve as environmental conditions shift over time. Our knowledge graph captures these temporal transitions by dynamically adjusting the edges between POIs across different time steps. This enables agents to:
\begin{itemize}
    \item \textbf{Predict Future States:} Temporal edges allow agents to forecast how the environment will evolve, helping them plan more effectively for long-term objectives.
    \item \textbf{Account for Dynamic Risks:} As hazards like water currents change over time, edge weights are updated to reflect these conditions, ensuring agents avoid paths that are likely to become dangerous.
\end{itemize}

\begin{figure}[htbp]
\centering
\includegraphics[width=\linewidth]{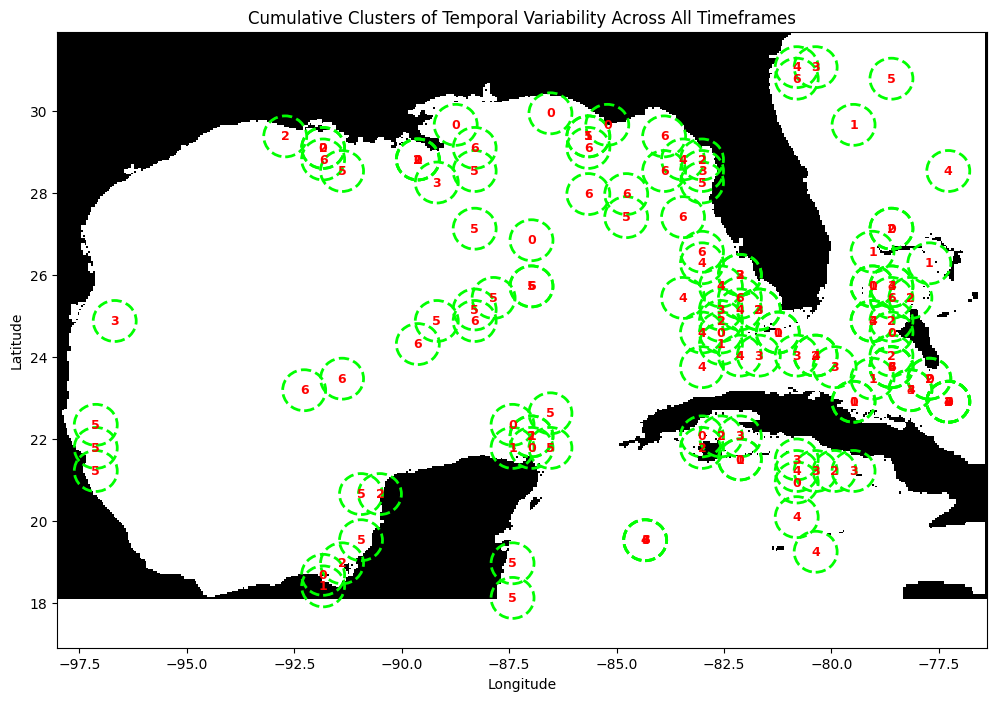}
\caption{Temporal POI cluster realizations across each timeframe.}
\label{fig:temporal-evolution}
\end{figure}

The temporal progression of the ROBUST Knowledge Network is illustrated in Figure \ref{fig:temporal-evolution}, which captures the cumulative clusters of POIs and temporal variability across all timeframes, and Figure \ref{fig:robust-sequence}, which shows network states at three specific time frames.

\begin{figure}[ht]
\centering
\begin{tikzpicture}
    \node[inner sep=0pt] (image) at (0,0) {\includegraphics[width=0.75\columnwidth]{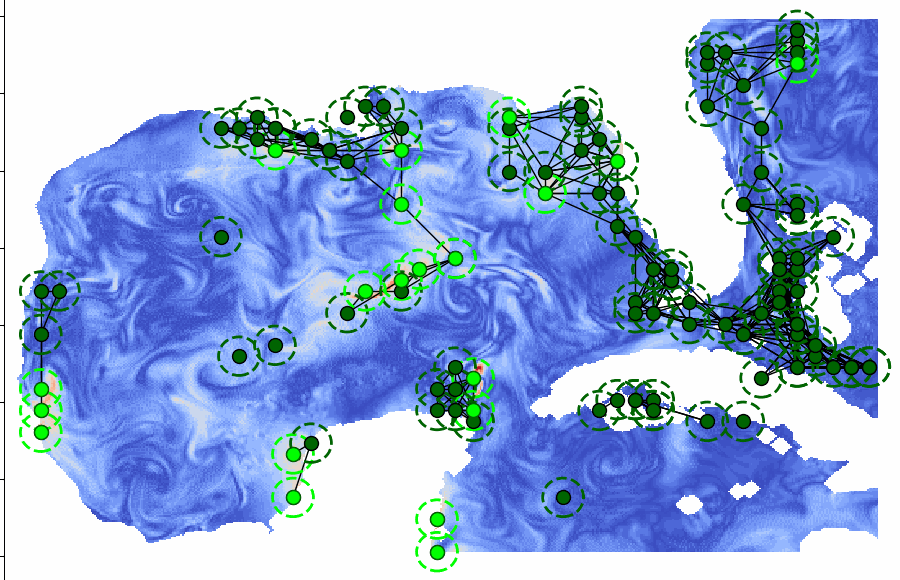}};
    \draw[black] (image.south west) rectangle (image.north east); 
\end{tikzpicture}
\caption{TED predictor shows node activations for frame 1}
\label{fig:ted-predictor}
\end{figure}

\subsection{Path Evaluation}

We evaluate the performance of WAITR using three key metrics: POI coverage, event counts (capturing environmental changes), and hazard avoidance. These metrics assess the algorithm's ability to guide agents toward significant events, adapt to dynamic conditions, and minimize exposure to hazards.

\section{Experimental Setup}

To evaluate the performance of the proposed WAITR algorithm, we conducted simulations using a case study based on the Gulf of Mexico (GoM). This section describes the experimental setup, including the data sources, parameter settings, definition of POIs and hazards, and the computational environment. 

\subsection{Data Source and Parameters}

Environmental data for the GoM was obtained from the HYCOM (Hybrid Coordinate Ocean Model) provided by the Naval Research Laboratory (NRL) \cite{chassignet2007hycom}. This model provides real-time updates of various oceanographic parameters at a resolution of 1/25° (GOMl0.04), including:

\begin{itemize}
\item Sea water temperature
\item Salinity
\item Current velocities
\item Sea surface height
\end{itemize}

\vspace{1mm}
The simulation used a planning horizon \(T\) of \textbf{6 time steps}, a discount factor \(\gamma\) of \textbf{0.9}, and an observational radius (\(circle\_radius\)) of \textbf{0.5 degrees}.  This observational radius was arbitrarily selected for the experimental setup. The justifications for the planning horizon and discount factor are as follows:

\begin{itemize}
\item \textbf{Planning Horizon (\(T\)):} A planning horizon of 6 time steps was selected to balance short-term gains with long-term optimization. In a highly dynamic environment like the GoM, predicting conditions too far into the future can be unreliable. 6 time steps allow for a reasonable lookahead without over-reliance on uncertain forecasts.
\item \textbf{Discount Factor (\(\gamma\)):} A discount factor of 0.9 strikes a balance between prioritizing immediate rewards and future opportunities. A value closer to 1 would place more emphasis on long-term gains, while a value closer to 0 would prioritize immediate events. This choice allows the algorithm to consider both present and future rewards effectively.
\end{itemize}

\subsection{Points of Interest and Hazards}

POIs were defined as regions exhibiting significant temperature differentials between consecutive time frames, and a POI is considered covered if it falls within an agent's observational radius. This approach aims to identify dynamic areas where updated readings are most valuable. Hazards were defined as regions with strong UV currents that could potentially hinder the movement of AUVs. 

\subsection{Agents and Simulation Environment}

The simulations involved \textbf{three} AUVs tasked with collecting data in the GoM. Each AUV was assigned an initial starting location, determined by the top three waypoints identified by the WPR clustering algorithm in the first time step. This strategy ensures initial deployment to areas with high potential for significant events. Each AUV operated independently within its assigned pathlet. The simulations were conducted using Google Colab, leveraging GPU compute units for parallel processing to ensure scalability of the solution. 

\subsection{Evaluation Metrics}

The performance of the WAITR algorithm was evaluated using the following metrics:

\begin{itemize}
\item \textbf{POI Coverage:} The percentage of significant POIs visited by the AUVs.
\item \textbf{Event Counts:} The number of environmental changes (temperature differentials) captured by the AUVs. 
\item \textbf{Hazard Avoidance:} The extent to which AUVs avoided traversing hazardous regions (strong UV currents). 
\end{itemize}

\subsection{Baseline Comparison}

The performance of WAITR was compared against a traditional greedy algorithm that prioritizes visiting the nearest POI with the highest immediate reward at each time step. This comparison highlights the benefits of incorporating long-term planning and future predictions into the path planning process.

\subsection{Implementation of the Greedy Algorithm}

The baseline greedy algorithm used for comparison operates as follows:

\begin{enumerate}
\item At each time step, the algorithm determines the nearest POI with the highest reward (event count) within the AUV's observational range.
\item The AUV moves towards this POI. 
\item If multiple POIs have the same highest reward, the algorithm selects the one closest to the AUV's current location.
\item If no POIs are within the AUV's observational range, it remains at its current location.
\end{enumerate}

This greedy approach focuses solely on maximizing immediate rewards without considering future predictions or potential hazards.

\section{Results}

\subsection{Event Coverage Analysis}
The Event Coverage Ratio (ECR) is a metric developed to quantify how effectively significant environmental events were captured within the range of the AUVs’ sensors. As the results demonstrate, the WAITR planner achieved a higher ECR in almost all frames, particularly in the later stages of the mission. This is due to the planner’s ability to incorporate future reward potentials into its decision-making, allowing for more efficient long-term coverage of dynamic POIs.

\subsection{Waypoint Placement and Potential Coverage}

Before evaluating the path planning performance of the WAITR algorithm, we first analyze the potential coverage achievable based on initial waypoint placement. Given the limited observational range of the AUVs, strategically positioning waypoints is crucial to maximize the potential for capturing significant events.  Table \ref{tab:cluster_comparison} shows the percentage of POIs within the observational range of the top 1, 5, 10, and 20 waypoints identified using the Weighted Proximal Recurrence (WPR) clustering algorithm.

\begin{table}[h]
\centering
\resizebox{\columnwidth}{!}{%
\begin{tabular}{lcccc}
\hline
\textbf{Cluster Approach} & \textbf{Top 1} & \textbf{Top 5} & \textbf{Top 10} & \textbf{Top 20} \\ \hline
Aggregated WPR Counts                 & 1256           & 4726           & 7308            & 10378           \\
WPR\% (total=14273)        & 8.8\%          & 33.11\%        & 51.2\%          & 72.7\%          \\
\end{tabular}
}
\caption{Coverage comparisons using WPR clustering}
\label{tab:cluster_comparison}
\end{table}

These results demonstrate that even with a limited number of strategically placed waypoints, a significant portion of the POIs can be potentially covered. This highlights the importance of the WPR algorithm in identifying key locations for initial waypoint placement. 

\subsection{Event Coverage Ratio (ECR)}

The Event Coverage Ratio (ECR) measures the proportion of significant environmental events that are actually captured by the AUVs during their missions. This metric takes into account the paths generated by the WAITR algorithm and the agents' observational range. It is calculated as:

\[
ECR = \frac{\text{Number of Covered Events}}{\text{Total Number of Significant Events}}
\]

\begin{figure}[htbp]
\centering
\begin{subfigure}{.7\linewidth}
  \centering
  \includegraphics[width=\linewidth]{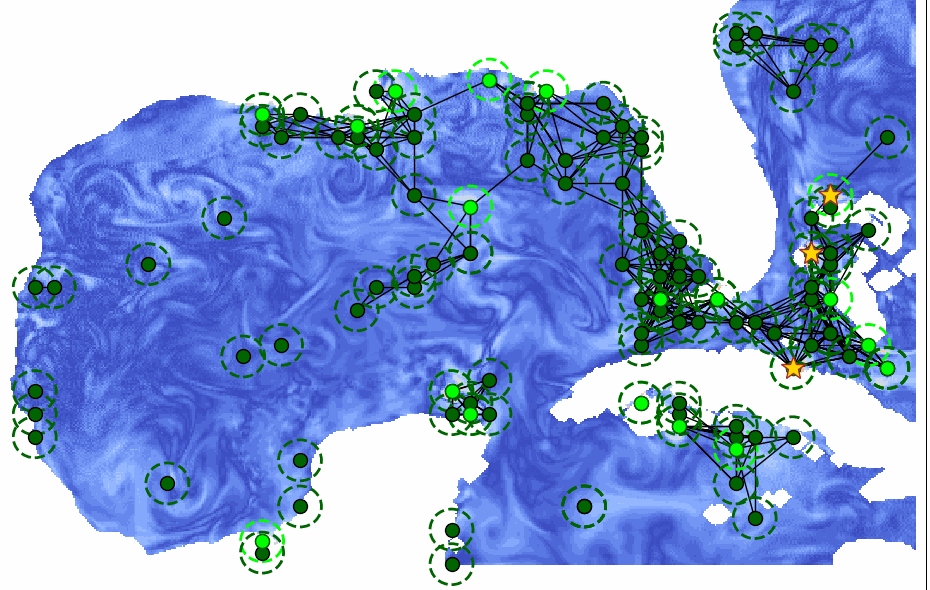}
  \caption{Frame 0}
\end{subfigure}

\begin{subfigure}{.7\linewidth}
  \centering
  \includegraphics[width=\linewidth]{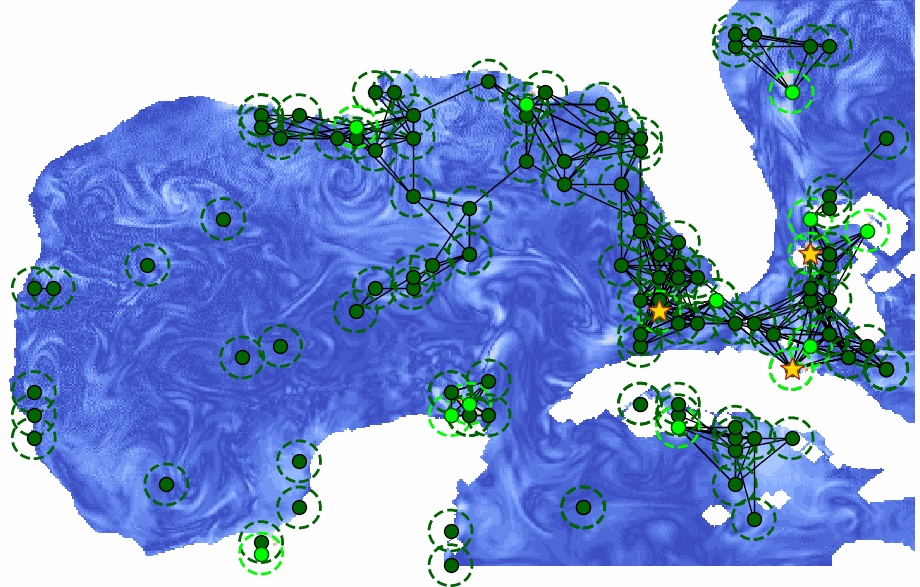}
  \caption{Frame 1}
\end{subfigure}

\begin{subfigure}{.7\linewidth}
  \centering
  \includegraphics[width=\linewidth]{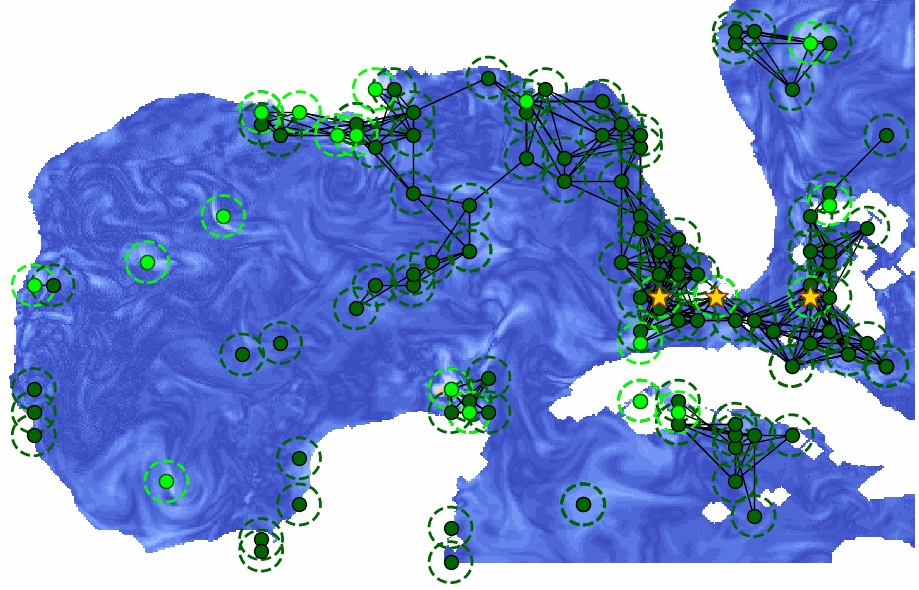}
  \caption{Frame 2}
\end{subfigure}

\caption{Sequential visualization of ROBUST-network path planning for three AUVs across Frames 0, 1, and 2.}
\label{fig:robust-sequence}
\end{figure}

The visualization in Figure \ref{fig:robust-sequence} shows the temporal variability in path planning, where green nodes represent waypoints identified by WPR clustering, and lime green nodes indicate active waypoints based on temporal significance and sensor coverage. Golden stars represent optimal sensor locations as determined by the WAITR strategy.

\subsection{Path Planning Comparison: WAITR vs. Greedy Algorithm}
To assess the efficiency of multi-agent spatiotemporal path planning strategies, we compared the performance of the WAITR (Weighted Aggregate Inter-Temporal Reward) planner with a traditional greedy algorithm. As shown in Table \ref{tab:glider_efficiency}, the WAITR planner consistently outperformed the greedy planner, achieving a total coverage of 27.1\%, compared to the greedy planner’s 23.56\%. Furthermore, the WAITR planner demonstrated improved performance in later timesteps, where the greedy planner’s efficiency declined sharply. This suggests that the WAITR planner is better suited for dynamic environments where future conditions must be anticipated.

\begin{figure}[htbp]
\centering
\begin{subfigure}{.52\linewidth} 
  \centering
  \includegraphics[width=\linewidth]{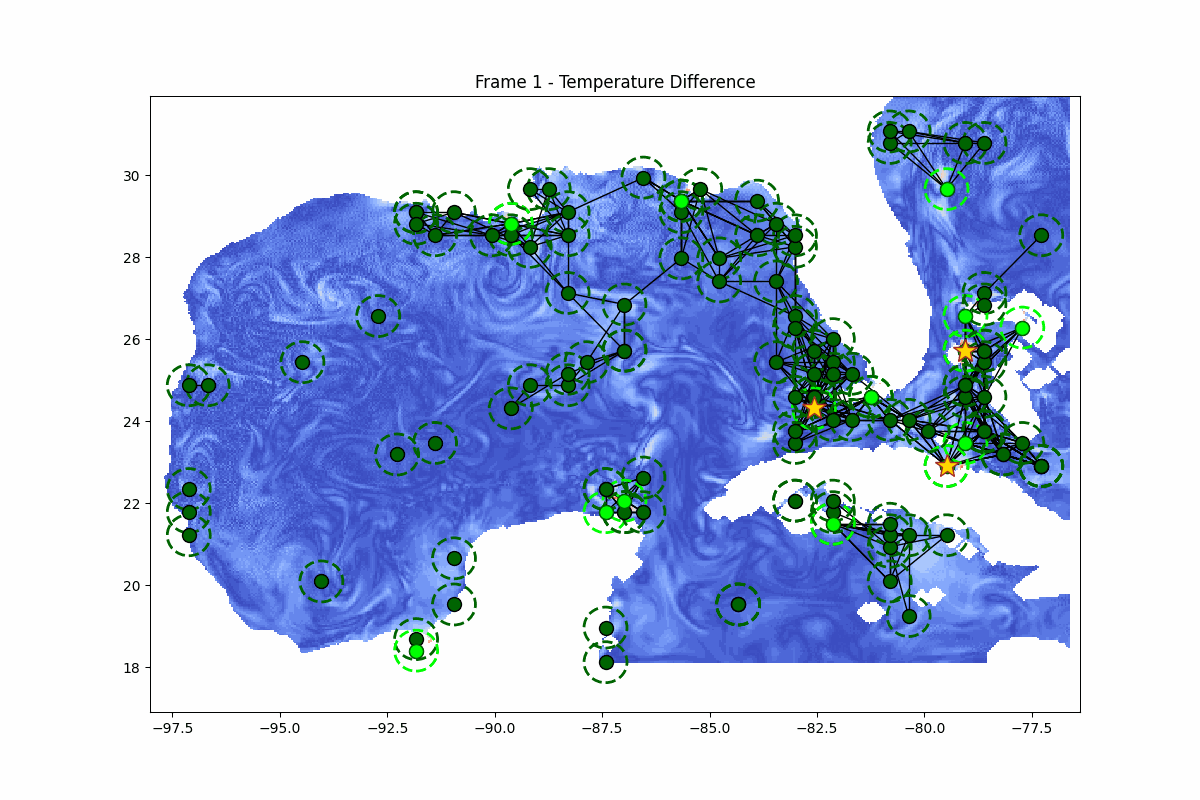}
  \caption{Frame 0}
\end{subfigure}
\hspace{-10mm} 
\begin{subfigure}{.52\linewidth} 
  \centering
  \includegraphics[width=\linewidth]{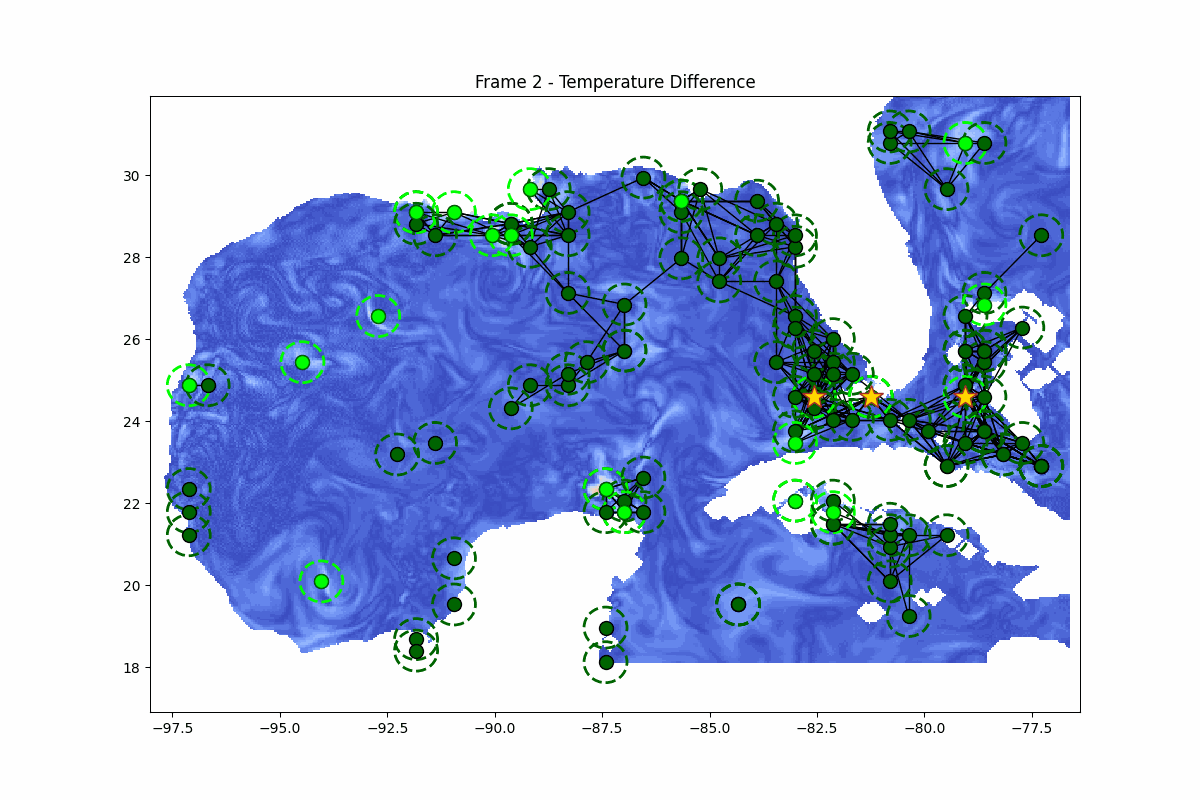}
  \caption{Frame 1}
\end{subfigure}
\vspace{2mm} 
\begin{subfigure}{.52\linewidth} 
  \centering
  \includegraphics[width=\linewidth]{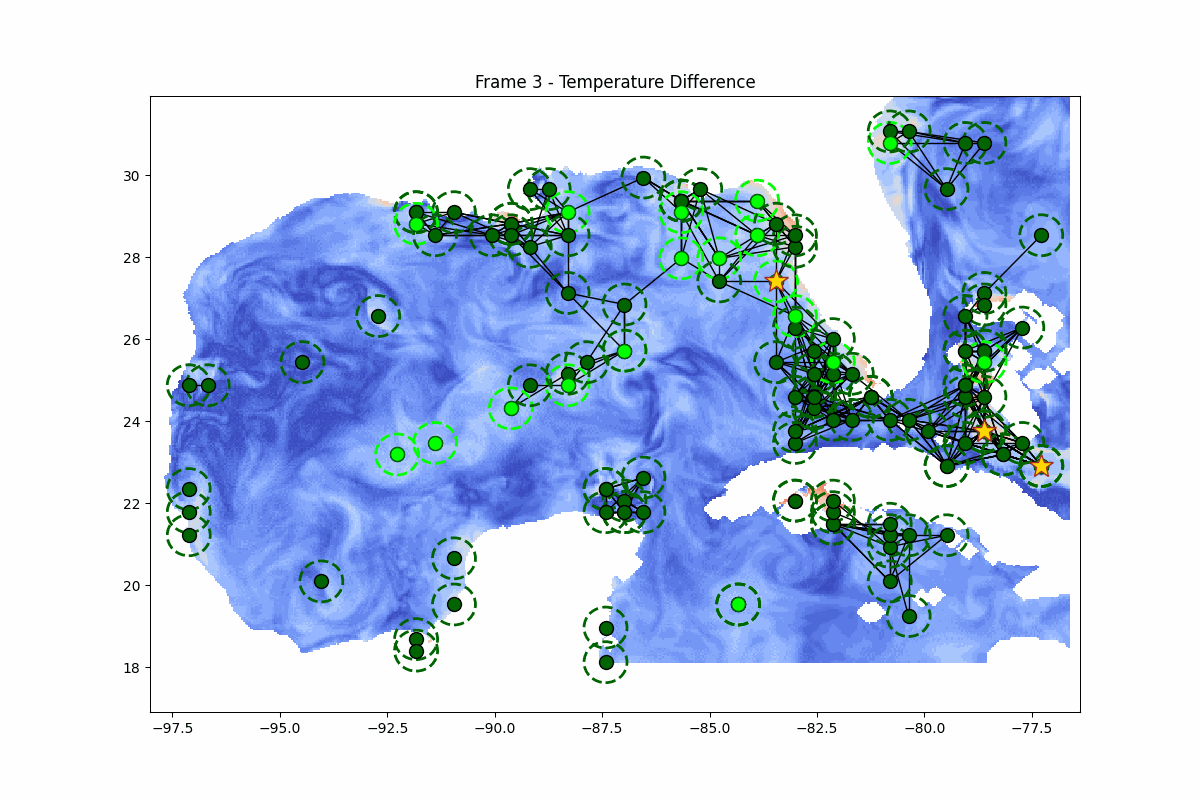}
  \caption{Frame 2}
\end{subfigure}
\hspace{-10mm} 
\begin{subfigure}{.52\linewidth} 
  \centering
  \includegraphics[width=\linewidth]{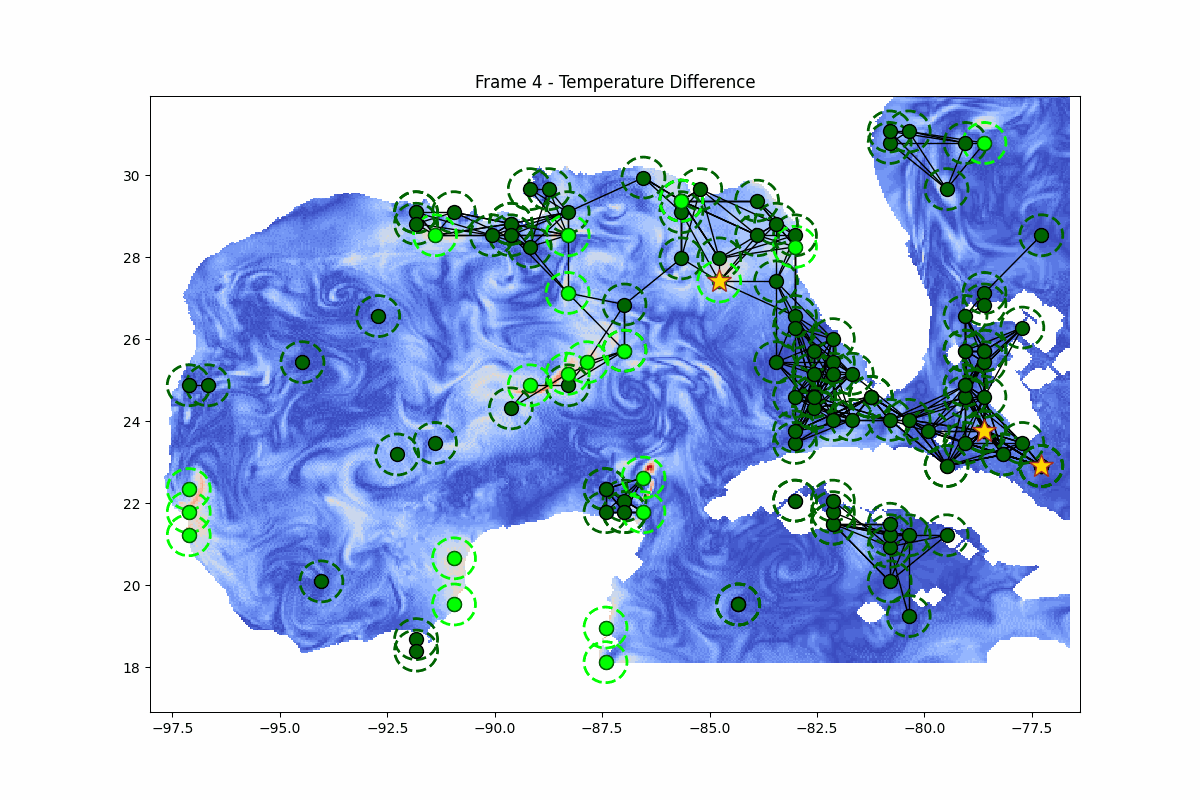}
  \caption{Frame 3}
\end{subfigure}
\vspace{2mm} 
\begin{subfigure}{.52\linewidth} 
  \centering
  \includegraphics[width=\linewidth]{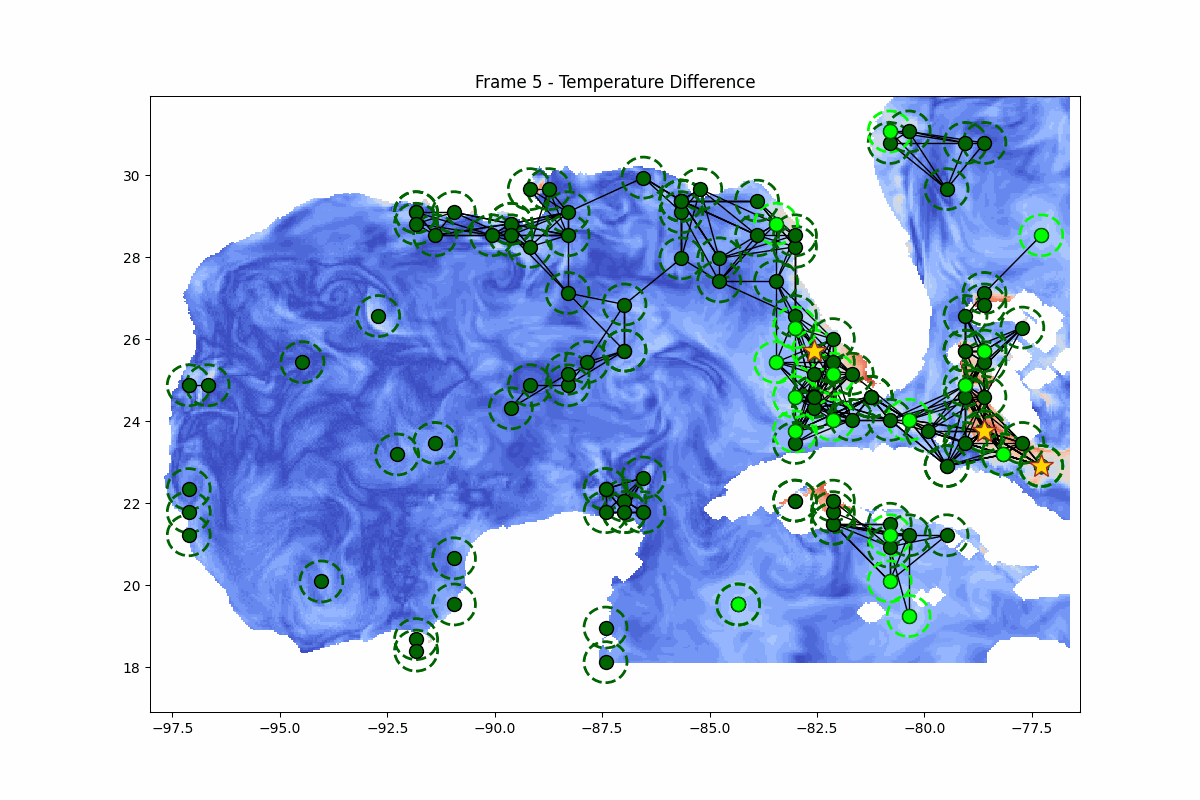}
  \caption{Frame 4}
\end{subfigure}
\hspace{-10mm} 
\begin{subfigure}{.52\linewidth} 
  \centering
  \includegraphics[width=\linewidth]{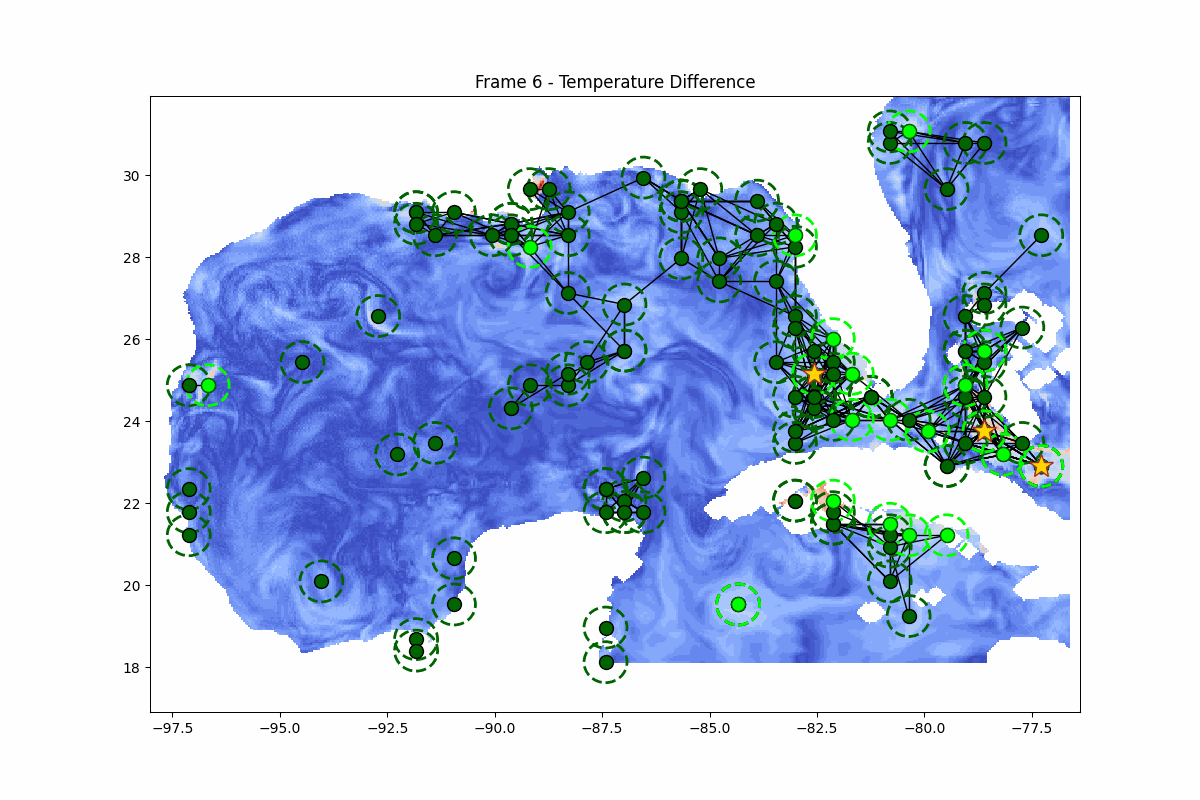}
  \caption{Frame 5}
\end{subfigure}
\caption{Sequential visualization of path planning across Frames 0 to 5, comparing \textbf{WAITR} (agent positions marked by stars). Dark green circles represent inactive waypoints for that timeframe, while lime green circles indicate active waypoints. The dotted line indicates the agent's observational radius. }
\label{fig:robust-sequence-all}
\end{figure}

Figure \ref{fig:robust-sequence-all} visualizes the paths generated by the WAITR algorithm across six time frames. The agent positions, marked by stars, demonstrate how WAITR dynamically adjusts the AUVs' trajectories to prioritize areas with high event counts (represented by lime green circles).  In contrast, the greedy algorithm would assign each agent to a single waypoint based on the highest initial event count. It would then remain at that waypoint until the event's value is depleted, even if more promising opportunities emerge elsewhere.  This myopic strategy leads to static positioning and potentially missed opportunities as event values change over time. WAITR, on the other hand, adapts to dynamic conditions by reevaluating paths and repositioning agents to maximize long-term event coverage. For example, in Frame 3, we observe that agents have moved away from their initial waypoints in Frame 0, repositioning themselves to capture newly emerging events with higher values.

\begin{table}[h]
\centering
\resizebox{\columnwidth}{!}{%
\begin{tabular}{lcc}
\hline
\textbf{Timestep} & \textbf{WAITR Planner (\%)} & \textbf{Greedy Planner (\%)} \\ \hline
Frame 0                   & \textcolor{red}{476}                        & \textcolor{green!60!black}{1463}                         \\
Frame 1                  & \textcolor{red}{18}                      & \textcolor{green!60!black}{116}                        \\
Frame 2                  & \textcolor{green!60!black}{1236}                       & \textcolor{red}{674}                        \\
Frame 3                 & \textcolor{green!60!black}{14}                      & \textcolor{red}{5}                         \\ 
Frame 4                 & \textcolor{green!60!black}{305}                       & \textcolor{red}{0}                         \\ 
Frame 5                 & \textcolor{green!60!black}{430}                       & \textcolor{red}{0}                         \\ 
Frame 6                 & \textcolor{green!60!black}{332}                       & \textcolor{red}{187}                         \\ \hline
\textbf{Total (10378)}                 & \textbf{2811 (27.1\%)}                       & \textbf{ 2445 (23.56\%)}                         \\ \hline
\end{tabular}
}
\caption{Efficiency comparison of WAITR and Greedy planning strategies for different numbers of gliders.}
\label{tab:glider_efficiency}
\end{table}

As shown in Table \ref{tab:glider_efficiency}, WAITR consistently outperforms the greedy algorithm in terms of POI coverage, achieving a higher Event Coverage Ratio (ECR) in most timeframes and overall.

\section{Discussion}

\subsection{Comparison of WAITR with Greedy Algorithms}
The results show that the WAITR algorithm significantly outperforms greedy algorithms in dynamic environments. Greedy algorithms, which make decisions based on immediate gains, fail to account for future environmental changes or long-term optimization. In contrast, WAITR integrates future temporal and spatial rewards into its decision-making process, allowing for more robust long-term planning.

The coverage metrics clearly demonstrate that WAITR achieves greater efficiency in covering critical Points of Interest (POIs) compared to the greedy algorithm, particularly in later timesteps. WAITR's adaptability to these changing conditions ensures that it remains effective even as hazards emerge or POIs shift in importance.

\subsection{Contribution of Knowledge Graph Integration}
The integration of knowledge graphs into WAITR is a key factor in its enhanced performance. Knowledge graphs dynamically update to reflect current and predicted environmental conditions, enabling agents to query the graph for both immediate and long-term path planning.

By encoding spatial and temporal relationships within the graph, WAITR agents can foresee shifts in environmental hazards  and adjust their paths accordingly. The knowledge graph's ability to maintain up-to-date information about the environment and potential hazards allows WAITR to outperform static, non-graph-based approaches by incorporating this real-time data into decision-making.

\subsection{Pathlet-Based Scalability and Performance}
The use of pathlets in WAITR contributes significantly to its scalability. By dividing the environment into smaller, more manageable subgraphs, WAITR reduces the computational complexity of path planning. This localized decision-making approach allows agents to focus on specific areas (pathlets) without overwhelming the system with global computations.

The pathlets' localized optimization also ensures that agents can adapt to immediate environmental changes while still considering broader global conditions. Additionally, the use of lookup tables for precomputed shortest paths within each pathlet further reduces the computational burden, allowing for efficient real-time path planning even in complex environments.

\subsection{Multi-Agent Coordination and Hazard Mitigation}
WAITR demonstrates strong multi-agent coordination capabilities by utilizing the knowledge graph to avoid conflicts between agents. The graph stores data about other agents' paths and hazards, allowing WAITR to direct agents to different POIs or less hazardous regions to ensure comprehensive coverage while minimizing the risk of overlap.

The hazard mitigation strategies employed by WAITR are particularly effective. Agents can assess the real-time hazard levels, encoded as edge weights in the knowledge graph, and avoid risky regions. This capability ensures that agents navigate safely while still fulfilling their data collection objectives in dynamic environments.

\subsection{Cumulative Scoring and Future Predictions}
A key strength of WAITR is its cumulative scoring system, which incorporates future potential rewards into the path planning process. This forward-looking strategy allows agents to make decisions not just based on immediate POI importance but also on predicted changes in environmental conditions and POI value over time.

The results demonstrate that WAITR consistently achieves better long-term outcomes compared to greedy algorithms. By factoring in future transitions encoded in the knowledge graph, WAITR ensures that agents are better positioned to avoid future hazards and capture high-value data points that may emerge later in the mission.

\subsection{Real-Time Adaptation and Knowledge Graph Updates}
WAITR's ability to adapt in real time is one of its most notable advantages. The knowledge graph is updated dynamically as new data is collected, allowing agents to adjust their paths based on the latest environmental information. This ensures that WAITR remains responsive to emerging hazards or changing POI priorities, leading to safer and more effective path planning.

The real-time updates enable agents to continuously refine their strategies, ensuring optimal coverage and hazard avoidance as conditions evolve. However, this real-time adaptability comes with challenges, such as the need for efficient graph updates to minimize computational overhead.

\subsection{Limitations and Broader Applicability}

While the WAITR algorithm shows promising results, it's essential to acknowledge its limitations. One challenge lies in the computational overhead associated with real-time knowledge graph updates. As the number of agents or the complexity of the environment increases, maintaining real-time adaptability might require additional computational resources or lead to processing delays.

Another limitation stems from the accuracy of future environmental predictions. In highly dynamic or unpredictable environments, even sophisticated predictive models might struggle to provide accurate forecasts. While WAITR incorporates uncertainty through confidence scores and discount factors, inaccurate predictions can still impact the optimality of the generated paths.

Despite these limitations, the WAITR algorithm offers broader applicability beyond AUVs in the Gulf of Mexico. Its core principles of integrating a knowledge graph, pathlet-based planning, and cumulative scoring can be extended to various domains where dynamic path planning is crucial. For instance:

\begin{itemize}
    \item \textbf{Autonomous Surveillance and Patrolling:} WAITR can be applied to scenarios where agents need to survey an area, prioritizing regions with high likelihoods of events or recent changes, while considering potential risks and optimizing coverage over time.
    \item \textbf{Resource Exploration and Monitoring:} In domains like mining or environmental monitoring, WAITR can guide agents to explore and monitor areas with high resource potential, adapting to dynamic changes in resource distribution and environmental conditions.
    \item \textbf{Dynamic Traffic Routing:} WAITR's ability to predict future changes and optimize for long-term rewards can be valuable for routing vehicles in dynamic traffic conditions, minimizing congestion and travel time while considering potential hazards or road closures.
\end{itemize}

The scalability of the WAITR algorithm stems from the use of the PREP Mapper, which allows for prioritized waypoint placement based on POI likelihood occurrences. This targeted approach focuses computational resources on areas of higher interest, making the algorithm suitable for larger and more complex environments.

Future research can explore further enhancements to the WAITR algorithm, such as incorporating more sophisticated predictive models, developing strategies for multi-agent coordination within pathlets, and investigating efficient methods for real-time knowledge graph updates.

\section{Conclusion}
This paper has demonstrated the efficacy of the WAITR (Weighted Aggregate Inter-Temporal Reward) algorithm in dynamic, multi-agent environments, highlighting its significant advantages over traditional greedy algorithms. Through the integration of a knowledge graph for real-time updates and a cumulative scoring system, WAITR has shown marked improvements in both short-term and long-term decision-making. This facilitates efficient coverage of critical Points of Interest (POIs) while minimizing risks from hazardous environmental conditions.

The use of pathlets has proven effective for scalable, localized decision-making, enhancing system performance without sacrificing the global view. Additionally, dynamic updates to the knowledge graph have enabled agents to adapt seamlessly to changing conditions, demonstrating the algorithm’s robustness in environments like the Gulf of Mexico.

The success of WAITR in marine environments underscores its potential for adaptation to other dynamic settings, such as terrestrial and aerial environments. Future research could explore the integration of more advanced machine learning models for environmental condition predictions and enhanced multi-agent coordination. As autonomous systems continue to grow in complexity and scope, the WAITR framework offers a promising foundation for more intelligent, adaptive, and efficient path-planning strategies.

\section*{Acknowledgment}
This work was partly supported by the U.S. Department
of the Navy, Office of Naval Research (ONR), and Naval Research Laboratory under contracts N0073-16-2-C902 and N00173-20-2-C007, respectively.

\bibliographystyle{IEEEtran}
\bibliography{references}

\end{document}